\documentclass[
   ,final            
  ]
  {aipproc}

\layoutstyle{6x9}

\begin{document}

\title{Future Investigations of the Flavor Dependence of Sea Quark Helicities 
at STAR}


\author{J. Sowinski \it{for the STAR Collaboration}}{
  address={Indiana University Cyclotron Facility, 2401 Milo B. Sampson Ln.,
Bloomington IN 47408
sowinski@indiana.edu}
}

\begin{abstract}
The flavor dependence of polarized and unpolarized quark distributions
in the nucleon can lead to insights into the formation of the sea.
Drell-Yan measurements have pointed to flavor asymmetries in the
unpolarized distributions.  Collisions at $\sqrt{s}$~=~500 GeV with
polarized protons at RHIC will soon allow investigations of the flavor
separated polarized quark distributions via W production to complement
measurements from semi-inclusive DIS.  We report on STAR's current
plans, tracking upgrade, and expected sensitivities.

\end{abstract}

\maketitle


\section{Introduction}

~~~The identity of the nucleon comes from the valence quarks, 2 up and a down
for a proton, 2 down and an up for neutron.
But high energy scattering processes such as deep inelastic scattering (DIS)
and hadronic reactions, e.g. Drell-Yan (DY) and jet production, reveal that,
in addition, there is a sea of quarks, anti-quarks and gluons residing in
the nucleon.  Moreover the distributions of these change with the scale or
momentum transfer of the scattering.  Much of this conference is devoted to
reports on understanding how the spin of the nucleon is assembled from
the intrinsic spin and orbital motion of these partons.  One important
aspect of this quest is separating how the various flavors of the 
quarks and anti-quarks differ in their contributions and exploring whether
something more might be learned beyond the distributions themselves.

Simplified pictures of the origin of anti-quarks in the nucleon give
dramatically different expectations for the ratios of $\bar{u}$ to
$\bar{d}$ quarks in a proton.  In perturbative QCD they arise in
quark/anti-quark pair creation and, since the $\bar{u}$ and $\bar{d}$
masses are small, one would expect roughly equal numbers.  On the
other hand, if one considers the long range nuclear force mediated by
pions, the proton has a component that is made of a neutron and a
${\pi}^+$ resulting in an expected excess of $\bar{d}$ quarks. (For a
general overview see Ref. \cite{Garvey:2001yq}.) The E886 DY experiment
by the NuSea collaboration has generated significant interest by
demonstrating clear asymmetries in the momentum fraction (x) dependent
distributions of $\bar{u}$ and $\bar{d}$ \cite{Towell:2001nh}.
Corrections to pQCD models for generating the sea accounting for Pauli
blocking effects have not been able to explain the data.  A number of
models, some related to quark bag models, and some chiral symmetry
motivated are consistent with the unpolarized data.  However they tend
to make different predictions for the helicity dependence of the
flavor asymmetry, with chiral quark soliton models predicting larger
asymmetries for the polarized distributions than the bag 
models\cite{Barone:2003jp}.

In a recently published global analysis\cite{deFlorian:2008mr}, the
first to include both DIS and RHIC data, the most favored
distributions demonstrate significant differences between the
$\bar{u}$ and $\bar{d}$ helicity dependent distributions denoted by
$\Delta\bar{u}$ and $\Delta\bar{d}$.  However the uncertainties
overlap and symmetric distributions are still allowed.  In the
following I will describe how measurements of W production in
polarized proton-proton collisions at RHIC can provide further
constraints on these distributions focusing in particular on the
proposed measurements and techniques in the STAR detector.
\hskip 6pt

\section{W production at STAR}
\hskip 6pt

~~~While sensitivity to flavor separated distributions in DIS is
primarily through coincident detection of hadrons in the exit channel
\cite{Airapetian:2004zf,Alekseev:2007vi}, 
the sensitivity in polarized proton-proton collisions will
be attained via W production carried out at $\sqrt{s}$~=~500~GeV in RHIC.  
The production mechanism is primarily through the 
channel where a u quark combines with a $\bar{d}$
anti-quark to produce a W$^+$ and likewise a d with a $\bar{u}$ to 
produce a W$^-$.  In addition, the V-A coupling of the weak interaction
allows only left-handed quarks to combine with right-handed anti-quarks 
providing a maximal parity violating signal and thus the necessary 
spin determination.  Finally, the handedness of the neutrinos in the 
W decay results in a partial transfer of the initial state kinematics, q
and $\bar{q}$ relative momentum fractions, into the final state 
decay products.  The required measurements consist of determining A$_L$,
the change in cross section with a single beam spin helicity flip, for 
W$^+$ and W$^-$ for flip of each beam independently.  The two W$^+$ 
measurements
provide sensitivity to two linear combinations of $\Delta$u(x) and 
$\Delta\bar{d}$(x) and 
the two W$^-$ measurements to $\Delta$d(x) and $\Delta\bar{u}$(x).  At 
forward angles for
W$^-$ the measurements separate almost completely into the two pure 
distributions.

At STAR we will focus on the W decay mode producing electrons or
positrons, W$^+ \rightarrow e^+ + \nu$ or W$^- \rightarrow e^- +
\bar{\nu}$.  The momentum of the high p$_T$ leptons will be measured
with electromagnetic calorimeters and the crucial charge sign with tracking
in a magnetic field.  The STAR detector\cite{Ackermann:2002ad} 
covers a full 2$\pi$ in
azimuth and is embedded in a solenoidal magnet with field strength of
0.5 T.  A major portion of the internal volume is occupied by a large
time projection chamber (TPC) to track charged particles over a radius from
0.6m to 1.9m.  The existing tracking should be sufficient for pseudo-rapidity
(scattering angle) $\mid\eta\mid < 1$ ($\theta > 37 ^{\circ}$).  Outside
of the TPC reside two Pb/scintillator sampling electromagnetic calorimeters
with the barrel covering -1 $< \eta < $1 and the endcap covering  
1.1 $< \eta < $2 (38$^{\circ} > \theta > 15^{\circ}$).  The calorimeters
each have longitudinal segmentation to help in providing background 
rejection as will be discussed below.  These include separate read out of
the first few layers of the calorimeters to provide a pre-shower signal and
fine position resolution shower maximum detectors to measure the transverse
shower profile.   In addition, the endcap calorimeter has a separately 
read out final layer to provide a post-shower signal.

To increase the reliable tracking range to $\eta \sim $2 a Forward GEM Tracking
upgrade\cite{Simon:2008qk} has been proposed and funded.  Six GEM chambers 
transverse to the 
beam will be
placed between the TPC and the beam in line with the endcap calorimeter.
The $<$80$\mu$m resolution of these planes along with a transverse beam 
constraint, points from the TPC where available, and the hit in the calorimeter
shower max detector should provide sufficient charge sign separation for
the 20 GeV/c$<$p$_T<$40 GeV/c electrons and positrons of interest.

\section{Simulations and Projected Sensitivity}

~~~Extensive simulations have been carried out to verify that the W signal can
be separated from the voluminous hard pQCD hadronic backgrounds present and
then estimate realistic expected statistical errors on the measurements 
including remaining necessary background subtraction and projected integrated
luminosities.  Pythia, which agrees well with NLO calculations of W 
yields\cite{Nadolsky:2003ga},
was used with GEANT models of our detector for these studies.

\begin{figure}
\caption{Projected sensitivity for 300 pb$^{-1}$.  At left is shown the 
parity violating asymmetry for W$^+$ production as measured by positron
detection.  The asymmetry is plotted vs. the positron pseudo-rapidity.
The curves represent predictions at NLO based on a range of PDFs that 
fit current data.  At right are the same quantities for W$^-$ production
detected as electrons.
}
\includegraphics[height=.4\textheight]{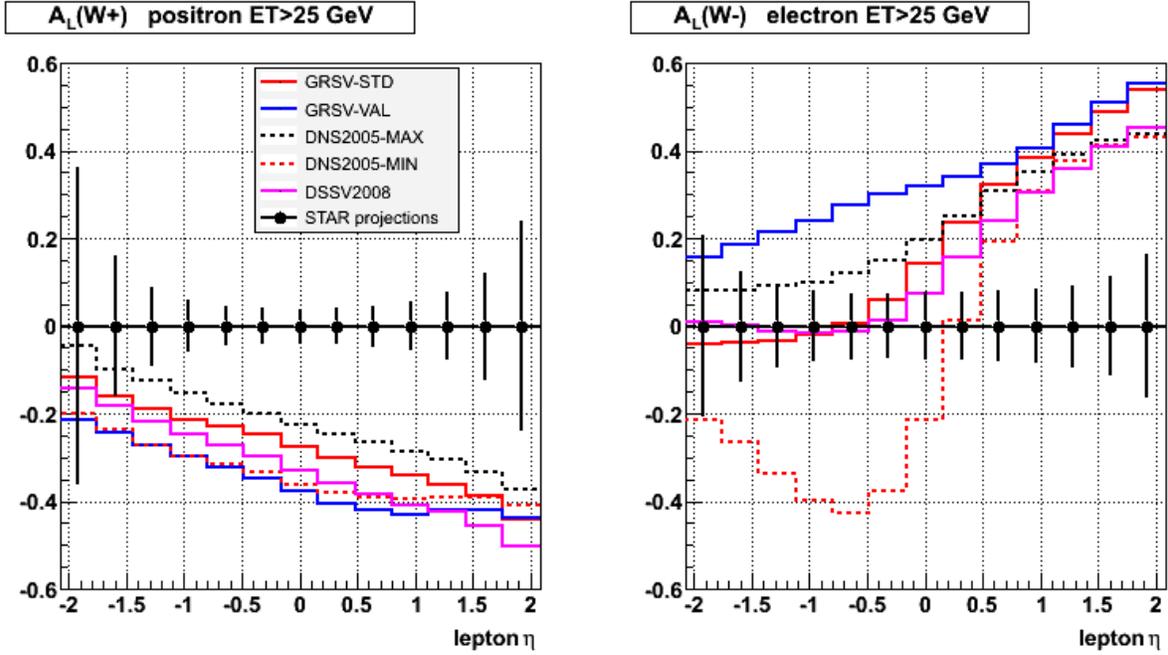}
\end{figure}

The signal for the channel of interest is an isolated electron with
p$_T >$ 20 GeV/c and a neutrino escaping undetected.  The primary
backgrounds are QCD jets which happen to have a particle that deposits
a similarly high p$_T$ in an electromagnetic calorimeter.  The STAR
detector is not sufficiently hermetic to reconstruct the neutrino with
missing transverse energy.  Thus we employ three classes of cuts to
reject background while preserving the signal.  The first is isolation
around the electron candidate.  Tracks and hits in the calorimeter
within r$<$0.45 in $\eta$-$\phi$ space are used to reject backgrounds
based on the accompanying jet particles.  The second class rejects
events with tracks and calorimeter energy opposite in $\phi$ from the
candidate, as most jets will have its partner of the dijet within the
acceptance.  Finally details of the candidate interaction in the
detector are used, such as the transverse profile at the shower
maximum detector and energy in the pre- and post-shower detectors.
The calorimeters are only one hadronic interaction length long so most
hadrons do not deposit a large fraction of their energy in the
calorimeter.  Mesons, such as $\pi^0$s, which decay
electromagnetically need to be rejected by a lack of hits in the
tracking system.  Up to 30\% conversions are allowed while still
giving the background rejection considered here and sets a requirement
on material in front of the tracking system.  Together these cuts
reduce the original background by up to a factor of 1000 while
preserving over 80\% of the signal and we find that for p$_T >$28 GeV/c
the signal to background is better than 1:1.

The projected sensitivity of our measurements for both electrons and
positrons with 300 pb$^{-1}$ of integrated luminosity is shown in
Fig. 1.  The asymmetries from the two beams have been combined into
one plot by assigning negative pseudo-rapidity to scattering into the
hemisphere from which the respective beam came.  Also shown are
projections for a range of recent parton distribution functions
consistent with current data.  For the W$^-$ projections one sees
significant sensitivity to the current range of uncertainty in
$\Delta\bar{u}(x)$ for $\eta < 0$.  For $\eta > 0$ the narrow range of
projections reflects the relatively well known valence $\Delta{d}(x)$
distribution.  For W$^+$ there is no clean separation between
$\Delta\bar{d}(x)$ and $\Delta{u}(x)$ but still significant
sensitivity for the central $\eta$ range which can be used in constraining
global fits to the parton distribution functions.

A run with polarized proton collisions at 500 GeV is planned to begin
early in 2009.  In this first run at that energy we project we could
collect as much as 10 pb$^{-1}$.  With polarization above 50\% this
should be sufficient to measure a statistically non-zero A$_L$ for 
the W$^+$ channel and put us on track for the full statistics measurement.


\begin{theacknowledgments}
This work was funded in part by grant NSF-0758018.

\end{theacknowledgments}



\bibliographystyle{aipproc}   

\bibliography{WprogProcFin}

\IfFileExists{\jobname.bbl}{}
 {\typeout{}
  \typeout{******************************************}
  \typeout{** Please run "bibtex \jobname" to optain}
  \typeout{** the bibliography and then re-run LaTeX}
  \typeout{** twice to fix the references!}
  \typeout{******************************************}
  \typeout{}
 }

\end{document}